\documentclass[aps,pra,floatfix,showpacs,preprint,onecolumn]{revtex4-2}
\usepackage{graphicx}
\usepackage{amsmath, amsfonts, amssymb, bm}
\usepackage{braket}

\usepackage{amstext}
\usepackage{mathrsfs}

\begin{document}
\title{Nonlinear Compton scattering in a frequency-modulated field}
\author{Antonino Di~Piazza}
\email{a.dipiazza@rochester.edu}
\affiliation{Department of Physics and Astronomy, University of Rochester, Rochester, New York 14627, USA}
\affiliation{Laboratory for Laser Energetics, University of Rochester, Rochester, New York 14623, USA}
\author{Kenan Qu}
\affiliation{Department of Astrophysical Sciences, Princeton University, Princeton, New Jersey 08544, USA}

\begin{abstract}
When an electron is accelerated, it emits radiation. In the relativistic quantum realm the elementary radiation process is the emission of a single photon, a process known as nonlinear Compton scattering in the case of an electron moving in the presence of a strong electromagnetic wave. This process is typically described within the Furry picture, where the electromagnetic wave is described as a classical background field and the electron-positron field is quantized in the presence of that background field. Equivalently one can quantize the electron-positron field in the vacuum but then the photon emission process is described as a transition from an initial state to a final state both featuring, apart from the electron (the initial state) and the electron and the photon (the final state), the same coherent state of photons appropriately related to the electromagnetic wave. Here, we consider a more general situation where the initial and the final state feature the same squeezed coherent state. Then, we specialize to the case where the coherent state corresponds to a plane-wave field and the mostly populated modes of the coherent state are also squeezed. We show that, when quantum fluctuations induced by the squeezing in the coherent field are negligible, a condition well satisfied at available squeezing levels, the squeezing effects effectively reduce to a frequency modulation of the plane-wave field corresponding to the coherent state. By means of numerical examples we show that at already available squeezing levels the emission spectrum of nonlinear Compton scattering and the total photon yield can be altered significantly. Analytical explanations of the main numerical results are also provided.
\end{abstract}

\maketitle

\section{Introduction}
Strong-field QED is the branch of QED concerning with processes which occur in the presence of intense background electromagnetic fields \cite{Landau_b_4_1982,Itzykson_b_1980}. 

By limiting to the lightest charged particles, electrons and positrons, with mass $m$ and charge $e$ and $-e$, respectively ($e<0$), the typical electromagnetic field scale characterizing strong-field QED is the so-called critical field of QED or Schwinger field:  $F_{cr}=m^2c^3/(\hbar|e|)=1.3\times 10^{16}\;\text{V/cm}=4.4\times 10^{13}\;\text{G}$. The vacuum in the presence of an electric field of the order of $F_{cr}$ is unstable under electron-positron pair production whereas the interaction energy of the electron/positron intrinsic magnetic moment and a magnetic field of the order of $F_{cr}$ is of the order of its rest energy. 

Another important quantity characterizing a background electromagnetic field is the length scale at which it varies significantly. The corresponding quantum length is given by $\lambda_C=\hbar/mc=3.9\times 10^{-11}\;\text{cm}$ and is known as (reduced) Compton wavelength \cite{Landau_b_4_1982,Itzykson_b_1980}. A photon with wavelength $2\pi\lambda_C$ would have an angular frequency $\omega_C=c/\lambda_C$ and an energy equal to the electron rest energy: $\hbar\omega_C=mc^2$.

Due to the Lorentz and the gauge invariance of the theory, when an electron interacts with a background electromagnetic field, the physical observables must depend on the field strength and wavelength via Lorentz- and gauge-invariant quantities. Thus, two parameters can be introduced to characterize the strong-field QED regime: $\chi_0=\sqrt{-(F_0^{\mu\nu}u_{\nu}/c)^2}/F_{cr}$ and $\eta_0=(k_0u)/\omega_C$ \cite{Landau_b_4_1982,Dittrich_b_1985,Fradkin_b_1991,Baier_b_1998}. Here, $F_0^{\mu\nu}$ indicates the amplitude of the electromagnetic field tensor, $k_0^{\mu}$ is the typical field wave four-vector, $u^{\mu}$ is the electron four-velocity all in the laboratory frame, and $(ab)$ denotes the four-dimensional scalar product between two four-vectors $a^{\mu}$ and $b^{\mu}$, with the assumed metric signature $(+1,-1,-1,-1)$. The condition $\chi_0\gtrsim 1$ ensures that quantum effects like the recoil in photon emission are important in that background electromagnetic field. On the other hand, the parameter $\eta_0$ controls the importance of recoil effects when an electron already interacts with a single photon of the background field. This is why another (related) parameter is employed in practice to characterize the strong-field QED regime and it controls whether nonlinear effects in the electromagnetic field amplitude, i.e., multiphoton effects are important in electromagnetic processes or not. This quantity depends on the spacetime structure of the electromagnetic field. Below, the case of a background laser field will be studied. In this case, this parameter is given by $\xi_0=|e|E_0/(m\omega_0 c)$, where $E_0$ is the electric field amplitude of the wave and $\omega_0$ its central angular frequency or its typical time-variation scale \cite{Ritus_1985,Di_Piazza_2012,Blackburn_2020,Gonoskov_2022,Fedotov_2023}. The condition $\xi_0\sim 1$ for nonlinear effects to become important corresponds for an optical laser to an intensity of about $10^{18}\;\text{W/cm$^2$}$, which is routinely achieved at several high-power laser facilities \cite{Apollon,CLF,CoReLS,ELI,ZEUS}, whereas future multi-petawatt lasers aim at values of $\xi_0$ exceeding $100$ \cite{NSF_OPAL,SEL,Vulcan_20-20,XCELS}. Note that in the ideal case of a plane wave it is $\xi_0=\chi_0/\eta_0$.

The present world record laser intensity is of the order of $10^{23}\;\text{W/cm$^2$}$ \cite{Yoon_2021} and it corresponds to a peak electric field about one thousand times smaller than $F_{cr}$. Thus, in order to enter the strong-field QED regime it is presently necessary to employ ultra-relativistic electron beams. Nowadays, laser-accelerated electron beams with energies of the order of a few GeV have been produced \cite{Gonsalves_2019} and this has allowed for testing experimentally the predictions of strong-field QED \cite{Cole_2018,Poder_2018,Mirzaie_2024,Los_2026}. As we have already hinted, in the strong-field QED regime, the laser radiation is so intense that it can be treated as a background, given electromagnetic field. Consequently, the effects of the laser field on the electrons and positrons can be taken into account exactly from the on-set by quantizing the electron-positron Dirac field in the presence of the laser field itself rather than in vacuum (Furry picture) \cite{Furry_1951,Landau_b_4_1982,Fradkin_b_1991}. The remaining interaction among electrons/positrons and the dynamical radiation field can still be accounted for perturbatively as its strength is typically determined by the fine-structure constant $\alpha=e^2/(4\pi\epsilon_0\hbar c)\approx 1/137$ \footnote{It has been conjectured that that interaction also becomes ``strong''  at values of the quantum nonlinearity parameter $\chi_0$ such that $\alpha\chi_0^{2/3}\ll 1$ \cite{Di_Piazza_2012,Fedotov_2023}.}. 

It is intuitively clear that the properties of strong-field QED processes and, in particular, of radiation, depend on the characteristics of the laser field such as its intensity, pulse shape, and polarization, which have then been exploited as control parameters~\cite{Di_Piazza_2012,Fedotov_2023}. Apart from these ``classical'' control parameters, recent theoretical works~\cite{Khalaf_2023, np_Even2023, TzurPRR2024, TheidelPRX2024} have started exploring the properties of radiation in quantum light states. In Ref. \cite{Di_Piazza_2026} we have presented a new approach to controlling quantum radiation by manipulating the quantum vacuum fluctuations of the emitted radiation field through squeezing. Specifically, we have investigated a quadrature-squeezed vacuum field~\cite{Walls_Sq1983, GSApra2012b}, where the quantum fluctuations of the electromagnetic field are reduced below the vacuum level in one phase (the squeezed quadrature) and correspondingly enhanced in the orthogonal phase (the anti-squeezed quadrature), in accordance with the uncertainty principle. The results show that squeezing can be effectively used as a quantum control parameter on the photon spectrum emitted via nonlinear Compton scattering.

Experimentally, strong squeezing up to $15$ dB has been achieved using cavity-assisted nonlinear crystals, significantly enhancing the sensitivity of gravitational wave detectors~\cite{PRL_15db_2016}. The achievable squeezing is primarily limited by detector noise and photon losses when coupling into interferometer cavities. However, for applications such as controlling quantum emission probabilities, where a cavity is not required, higher squeezing may be feasible. Recent studies have proposed novel schemes utilizing fully ionized plasmas, potentially enabling squeezing levels from $20$ dB~\cite{Qu_PRE_entangle24} to $40$ dB~\cite{Qu_2025}.

To date, most quantum squeezing experiments have been restricted to visible wavelengths, dictated by the operational range of conventional nonlinear crystals. However, a series of recent experiments have begun to explore the extension of squeezing into shorter wavelengths via high-harmonic generation~\cite{TheidelPRX2024, TzurPRR2024}. With ongoing developments in plasma optics, the realization of squeezed vacuum states in the x-ray regime is becoming increasingly plausible~\cite{Qu_PRE_entangle24, Qu_2025}.

In the present paper we continue the investigation started in Ref. \cite{Di_Piazza_2026} by considering nonlinear Compton scattering with the initial photon state being a squeezed coherent state. Unlike in Ref. \cite{Di_Piazza_2026}, however, we study the case where squeezing affects the modes of the coherent state and not of the emitted radiation. After reporting the general theory based on the description of the intense light as a highly populated squeezed coherent state, we apply it to the process radiation by an electron driven by a strong squeezed plane-wave field with arbitrary squeezing and frequency content. As we will indicate (see also Ref. \cite{Di_Piazza_2026}, currently available squeezing amplitudes allow for neglecting the quantum fluctuations of the coherent state due to the squeezing, such that the effect of the squeezing itself onto the intense coherent light can be described as a frequency modulation. We show both analytically and numerically that by combining available high-power laser and squeezing technology, it is possible to significantly alter the yield and the spectrum of the radiation emitted by an electron via the squeezing strength and angle.

Below, we use units with $\hbar=c=\epsilon_0=1$.

\section{Theoretical model}

The Lagrangian density of strong-field QED is given by \cite{Landau_b_4_1982,Fradkin_b_1991,Di_Piazza_2012,Fedotov_2023}
\begin{equation}
\mathcal{L}=\bar{\psi}(x)\{\gamma^{\mu}[i\partial_{\mu}-e\mathcal{A}_{\mu}(x)]-m\}\psi(x)-\frac{1}{4}F_{\mu\nu}(x)F^{\mu\nu}(x)-e\bar{\psi}(x)\gamma^{\mu}\psi(x)A_{\mu}(x),
\end{equation}
where $\psi(x)$ is the electron-positron Dirac field and $A^{\mu}(x)$ ($\mathcal{A}^{\mu}(x)$) is the radiation (background) electromagnetic field \cite{Di_Piazza_2012,Fedotov_2023}. The field $\bar{\psi}(x)=\psi^{\dag}(x)\gamma^0$ is the Dirac conjugate of $\psi(x)$, with $\gamma^{\mu}$ being the Dirac gamma matrices, and $F^{\mu\nu}(x)=\partial^{\mu}A^{\nu}(x)-\partial^{\nu}A^{\mu}(x)$ is the electromagnetic tensor of the radiation field. The effects of the background field are taken into account exactly in the calculations by quantizing the Dirac field in the presence of the background field (see, e.g., Refs. \cite{Landau_b_4_1982,Fradkin_b_1991,Di_Piazza_2012,Fedotov_2023}). This amounts to finding a basis of solutions of the Dirac equation
\begin{equation}
\{\gamma^{\mu}[i\partial_{\mu}-e\mathcal{A}_{\mu}(x)]-m\}\psi(x)=0
\end{equation}
corresponding to positive- and negative-energy states. Here, we implicitly assume that the Dirac equation in the external field $\mathcal{A}^{\mu}(x)$ admits such solutions and that the positive and negative energies are separated by a finite gap at all times. This essentially amount to working in the so-called Furry picture \cite{Furry_1951} and we refer the reader to Ref. \cite{Fradkin_b_1991} for a detailed discussion on the conditions on the background field allowing for this treatment. 

The electromagnetic field $A^{\mu}(x)$ is quantized as in vacuum, i.e.,
\begin{equation}
\label{A}
A^{\mu}(x)=\sum_{r=0}^3\int(d^3\bm{k})[a_r(\bm{k})e^{-i(kx)}e^{\mu}_r(\bm{k})+\text{H.c.}],
\end{equation}
where $(d^3\bm{k})=d^3\bm{k}/[(2\pi)^32\omega_k]$, with $\omega_k=|\bm{k}|$, where $a_r(\bm{k})$ and $a^{\dag}_r(\bm{k})$ are the operators of photon annihilation and creation in the state with four-momentum $k^{\mu}=(\omega_k,\bm{k})$ corresponding to the polarization four-vector $e^{\mu}_r(\bm{k})$, and where H.c. stands for Hermitian conjugate. The annihilation and creation operators fulfill the commutation relations $[a_r(\bm{k}),a^{\dag}_{r'}(\bm{k}')]=2\omega_k\delta_{r,r'}(2\pi)^3\delta(\bm{k}-\bm{k}')$, whereas all other possible commutators vanish. The physical photon states correspond to the transverse modes with $r=1,2$ and it is then convenient to introduce the indices $a=0,3$ and $j=1,2$.

The amplitude of an arbitrary process within the Furry picture is computed by working within the interaction representation with the interaction among electrons, positrons, and photons being represented by the last term in $\mathcal{L}$ and taken into account perturbatively. 

Now, if the background electromagnetic field is a solution of Maxwell's equations in vacuum, the four-vector potential $\mathcal{A}^{\mu}(x)$ can be written in the Lorenz gauge in the form 
\begin{equation}
\label{A_B}
\mathcal{A}^{\mu}(x)=\sum_{j=1}^2\int(d^3\bm{k})[b_j(\bm{k})e^{-i(kx)}e^{\mu}_j(\bm{k})+\text{c.c.}],
\end{equation}
where $b_j(\bm{k})$ are complex functions and c.c. stands for complex conjugate. It was shown in Ref. \cite{Fradkin_b_1991} that in this case the Furry-picture approach is equivalent to the conventional interaction picture with the Dirac field being quantized in vacuum and the interaction Lagrangian density being given by $-e\bar{\psi}(x)\gamma^{\mu}\psi(x)[A_{\mu}(x)+\mathcal{A}_{\mu}(x)]$ but with the initial and the final states containing, apart from electrons, positrons, and radiation photons, the coherent state of photons $\ket{B}=D(b)\ket{0}$, where
\begin{equation}
D(b)=\exp\left\{\sum_{j=1}^2\int(d^3\bm{k})[b_j(\bm{k})a^{\dag}_j(\bm{k})-b^*_j(\bm{k})a_j(\bm{k})]\right\}
\end{equation}
is the so-called displacement operator \cite{Mandel_b_2013,Agarwal_b_2013}. The equivalence is easy to be demonstrated starting from the basic property of the displacement operator $D^{\dag}(b)a_r(\bm{k})D(b)=a_r(\bm{k})+\sum_{j=1}^2\delta_{r,j}b_j(\bm{k})$, which implies that
$D^{\dag}(b)A^{\mu}(x)D(b)=A^{\mu}(x)+\mathcal{A}^{\mu}(x)$ and from the procedure outlined in Ref. \cite{Furry_1951}. 

In order to formulate strong-field QED in the presence of an intense coherent squeezed state, we analogously assume that, apart from the electrons, positrons, and photons described by the radiation field $A^{\mu}(x)$, both the initial and the final states of the system contain the squeezed coherent state $\ket{ZB}=S(z)D(b)\ket{0}$, where
\begin{equation}
S(z)=\exp\left\{\frac{1}{2}\sum_{j=1}^2\int(d^3\bm{k})[z^*_j(\bm{k})a^2_j(\bm{k})-z_j(\bm{k})a^{\dag\,2}_j(\bm{k})]\right\}
\end{equation}
is the so-called squeezing operator \cite{Mandel_b_2013,Agarwal_b_2013}, with $z_j(\bm{k})=\zeta_j(\bm{k})\exp(i\theta_j(\bm{k}))$ being complex functions. Analogously as in the case of a non-squeezed coherent state, the assumption here is that the main modes of the field are so highly populated that the depletion due to the quantum interactions is negligible, which underlies the treatment itself of the background field as a given field \cite{Landau_b_4_1982}. Note that in Ref. \cite{Di_Piazza_2026} we have considered the state $\ket{BZ}=D(b)S(z)\ket{0}$ and we have pointed out that the two cases are related to each other because of the the identity $D(b)S(z)=S(z)D(B)$ \cite{Mandel_b_2013}, with
\begin{equation}
\label{B}
B_j(\bm{k})=\cosh(\zeta_j(\bm{k}))b_j(\bm{k})+\sinh(\zeta_j(\bm{k}))e^{i\theta_j(\bm{k})}b^*_j(\bm{k}),
\end{equation}
which can be easily proven by using the basic property of the squeezing operator
\begin{equation}
\label{SdagaS}
S^{\dag}(b)a_r(\bm{k})S(b)=a_r(\bm{k})+\sum_{j=1}^2\delta_{r,j}\left\{[\cosh(\zeta_j(\bm{k}))-1]a_j(\bm{k})-\sinh(\zeta_j(\bm{k}))e^{i\theta_j(\bm{k})}a^{\dag}_j(\bm{k})\right\}.
\end{equation}
The formulation in terms of the state $\ket{ZB}$ is more convenient for the case that we want to investigate in the present paper.

Now, by using Eq. (\ref{SdagaS}), one can easily show that
\begin{equation}
D^{\dag}(b)S^{\dag}(z)A^{\mu}(x)S(z)D(b)=D^{\dag}(b)A^{\mu}_Z(x)D(b)=A^{\mu}_Z(x)+\mathcal{A}^{\mu}_Z(x),
\end{equation}
where
\begin{align}
\label{A_Z}
A^{\mu}_Z(x)&=\sum_{r=0}^3\int(d^3\bm{k})[a_r(\bm{k})E^{\mu}_r(x,\bm{k})+\text{H.c.}],\\
\label{A_ZB}
\mathcal{A}^{\mu}_Z(x)&=\sum_{j=1}^2\int(d^3\bm{k})[b_j(\bm{k})E^{\mu}_j(x,\bm{k})+\text{c.c.}],
\end{align}
with $E^{\mu}_{0,3}(x,\bm{k})=e^{-i(kx)}e^{\mu}_{0,3}(\bm{k})$ and
\begin{equation}
\label{E_j}
E^{\mu}_j(x,\bm{k})=\cosh(\zeta_j(\bm{k}))e^{-i(kx)}e^{\mu}_j(\bm{k})-\sinh(\zeta_j(\bm{k}))e^{-i\theta_j(\bm{k})}e^{i(kx)}e^{\mu\,*}_j(\bm{k}).
\end{equation}
Thus, by repeating the reasoning followed in the non-squeezed coherent case, we conclude that under the above assumptions strong-field QED in the presence of a strong squeezed coherent state can be formulated by quantizing the electromagnetic field according to Eq. (\ref{A_Z}) and the Dirac field in the presence of the background electromagnetic field $\mathcal{A}^{\mu}(x)=\mathcal{A}^{\mu}_Z(x)$. In other words, the presence of the squeezing can be accounted for within the external-field framework but by also modifying the photon polarization states according to Eq. (\ref{E_j}).

The above approach holds within the validity range of the Furry approach and under the assumption that the background electromagnetic field is a free field. Analogously as in Ref. \cite{Di_Piazza_2026}, we specialize our treatment to the case of a background plane-wave field, whose corresponding ``dressed'' electron states are known as Volkov states \cite{Volkov_1935,Landau_b_4_1982}. We assume that the plane wave propagates along the positive $z$ direction and that it is polarized along the $x$ direction, corresponding to $j=1$, i.e., $b_j(\bm{k})\to\delta_{j,1}(2\pi)^2\delta(k_x)\delta(k_y)H(k_z)b(k_z)$, where $H(\cdot)$ is the Heaviside step function (for the sake of notational simplicity, we use the same symbols of the various coefficients even though they are now functions of a single variable). Thus, we have that $\mathcal{A}^{\mu}_Z(\phi)=A_0f_Z(\phi)e^{\mu}_x$, where $A_0>0$ is a constant describing the amplitude of the field, where
\begin{equation}
\label{A_ZB_plane}
A_0f_Z(\phi)=\int_0^{\infty}\frac{d\omega}{2\pi}\frac{1}{2\omega}[b(\omega)E(\phi,\omega)+\text{c.c.}],
\end{equation}
with $\phi=t-z$ and
\begin{equation}
\label{E_plane}
E(\phi,\omega)=\cosh(\zeta(\omega))e^{-i\omega\phi}-\sinh(\zeta(\omega))e^{-i\theta(\omega)}e^{i\omega\phi},
\end{equation}
and where $e^{\mu}_x=(0,1,0,0)$. In Eq. (\ref{E_plane}) we set $\zeta(\omega)=\zeta_1(\omega\bm{z})$ and $\theta(\omega)=\theta_1(\omega\bm{z})$.

At this point the computation of the probability of nonlinear Compton scattering follows exactly the standard procedure, which can be found in the literature (see the reviews \cite{Di_Piazza_2012,Fedotov_2023} and the references therein), with two important changes: 1) the background field is $\mathcal{A}^{\mu}_Z(\phi)$ and depends on the squeezing; 2) the emitted photon ``state'' appearing in the transition amplitude due to the squeezing is $E^{*\,\mu}_j(x,\bm{k})$ rather than $\exp(i(kx))e^{*\,\mu}_j(\bm{k})$. We have seen in Ref. \cite{Di_Piazza_2026} how this second difference can be exploited to control nonlinear Compton scattering when the emitted photons have frequencies in the optical regime, where squeezing is customarily realized experimentally. For this to be the case, we have considered a strong terahertz background field such that the Doppler-shifted photons emitted by a relativistic electron had a frequency in the optical range. Here, we would like to consider the somewhat complementary situation, where the background laser field is optical and the squeezing function $\zeta_r(\bm{k})$ is sharply peaked around the central mode of the laser field. Since we will also assume that the initial electron is ultrarelativistic, the emitted modes will essentially not be squeezed. We can then set
\begin{equation}
\label{b}
b(\omega)=A_0\sqrt{2\pi}\omega_0\tau e^{-\frac{\tau^2(\omega-\omega_0)^2}{2}},
\end{equation}
which corresponds to a Gaussian beam with central angular frequency $\omega_0$ and full-width half maximum of the intensity $\tau_{\text{FWHM}}$ of $2\sqrt{\log(2)}\tau$. Below, we will consider a long pulse, with $\omega_0\tau\gg 1$. Also, we consider the standard Lorentzian shape 
\begin{equation}
\label{zeta}
\zeta(\omega)=\frac{\zeta_0}{1+(\omega-\omega_0)^2/\Gamma^2},
\end{equation}
with $\zeta_0$ and $\Gamma$ being two positive constants, for the squeezing function and the constant value $\theta(\omega)=\theta_0$ for the squeezing angle. We also introduce the notation $\varphi=\omega_0\phi$ and $\xi(\varphi)=-\xi_0df_Z(\varphi)/d\varphi$, with $\xi_0=|e|A_0/m$.

Now, available laser intensities $I_0\gtrsim 10^{20}\;\text{W/cm$^2$}$ at $\omega_0=1.55\;\text{eV}$, corresponding to a wavelength of $0.8\;\text{$\mu$m}$, allow for $\xi_0\gg 1$ and typical pulse durations $\tau\gtrsim 30\;\text{fs}$ are such that $\omega_0\tau\gg 1$. Moreover, available electron beams have energies $\varepsilon\gtrsim 5\;\text{GeV}$ such that $\varepsilon\gg m\xi_0$ \cite{Di_Piazza_2014,Di_Piazza_2015,Di_Piazza_2021}. Under these conditions an electron typically emits a relatively large number of photons when crossing such a laser field but it is barely deviated from its initial direction. In fact, at each emission the deviation angle due to recoil is of the order of $m\xi_0/\varepsilon\ll 1$ \cite{Di_Piazza_2012}. Thus, as a first study of this process, we wrote a numerical code based on a method which is equivalent to the kinetic approach presented in Refs. \cite{Neitz_2013,Neitz_2014} (see also \cite{Di_Piazza_2010}), where the electron dynamics is one-dimensional (the electron moves in the opposite direction of the plane wave) and solely described by its light-cone momentum $p_-=\varepsilon-p_z\approx 2\varepsilon$ (more general approaches including the effects of the transverse motion of the electrons have been developed in the literature, see the review \cite{Gonoskov_2022}). The electron propagates according to the classical Lorentz equation such that its light-cone momentum remains constant except when it randomly emits a photon and recoils by lowering its value of $p_-$ by the corresponding light-cone momentum $k_-$ of the photon. The numerical code also includes the possibility that a photon decays into an electron-positron pair although this process turned out to be completely negligible for the numerical parameters considered below. The known probabilities of photon emission (nonlinear Compton scattering) and of pair production (nonlinear Breit-Wheeler process) have been employed within the locally-constant field approximations, which is generally valid for $\xi_0\gg 1$ (see the reviews \cite{Di_Piazza_2012,Gonoskov_2022,Fedotov_2023}). The only difference brought about by the squeezing is that the background field is given by Eq. (\ref{A_ZB_plane}), with the function $E(\phi,\omega)$ given by Eq. (\ref{E_plane}) rather than by $\exp(-i\omega\phi)$. This shows that under the present conditions the effects of the squeezing can be interpreted as a frequency modulation of the background laser field. In fact, the energies of photons emitted by a 5-GeV electron in a strong laser with $\omega_0=1.55\;\text{eV}$ lay in the MeV-domain and the effects of the squeezing can be neglected in the final state. However, effects like the quantum fluctuations of the modes corresponding to the intense coherent state could be included in the present approach by computing the emission (and pair-production) probability at higher orders in $\alpha$. As we have discussed in Ref. \cite{Di_Piazza_2026}, if the squeezed optical modes would have a sufficiently large number of photons that the corresponding classical nonlinearity parameter would be of the order of unity of larger, such higher-order terms should be included. However, presently and soon available values of squeezing are well below this threshold.

\section{Numerical results}
We present now the results of numerical simulations in which we compare the emission spectra obtained in the case of a standard plane wave with a Gaussian envelope ($\zeta(\omega)=0$) with the squeezed plane wave. 

\begin{figure}
\begin{center}
\includegraphics[width=0.9\columnwidth]{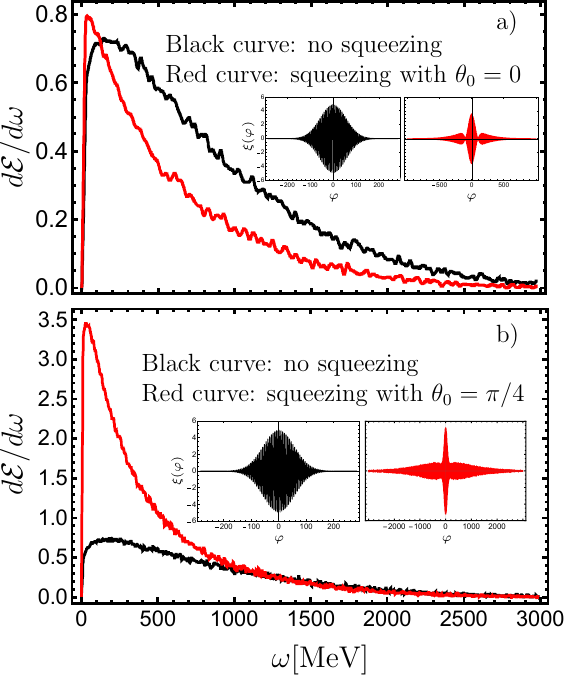}
\end{center}
\caption{Differential energy emitted per unit of photon energy in the presence of a plane wave with and without squeezing for numerical parameters given in the text. The insets show the respective functions $\xi(\varphi)$.}
\label{Spectra}
\end{figure}

We consider a laser field with central photon energy $\omega_0=1.55\;\text{eV}$, peak intensity $I_0=10^{20}\;\text{W/cm$^2$}$, corresponding to $\xi_0\approx 5$, and pulse duration $\tau_{\text{FWHM}}=40\;\text{fs}$. If the field were focused to a spot radius of $\sigma_0=3\;\text{$\mu$m}$, it would have an energy of $0.6\;\text{J}$. The squeezing parameters have been chosen as $\zeta_0=3.45$ (squeezing amplitude of about $30\;\text{dB}$) and $\Gamma=1.9\times 10^{-3}\;\text{eV}=0.9\;\text{THz}$. Finally, we simulated the collision of $10^5$ electrons counterpropagating with respect to the plane wave with a Gaussian energy distribution centered around $5\;\text{GeV}$, corresponding to $\chi_0\approx 0.3$, and with a standard deviation of $500\;\text{MeV}$. The emitted energy spectra $d\mathcal{E}/d\omega$ per unit of photon energy in the case of no squeezing and of squeezing, with $\theta=0$ and $\theta=\pi/4$ are plotted in Fig. \ref{Spectra}.a and Fig. \ref{Spectra}.b, respectively.
The figure clearly shows how it is possible to control the relative height of the spectrum by changing the squeezing angle $\theta_0$. In terms of total emitted energy $\mathcal{E}_e$ per electron, it is obtained that in the case without squeezing it is $\mathcal{E}_e\approx 803\;\text{MeV}$, whereas in the case $\theta_0=0$ ($\theta_0=\pi/4$) each electron emits on average $516\;\text{MeV}$ ($1600\;\text{MeV}$). Also, it turns out to be easier to enhance the emitted spectrum by increasing the squeezing angle $\theta_0$ than to suppress the spectrum for small values of $\theta_0$. This interesting feature can be understood qualitatively by looking at how the maximum value of the plane wave field $|f_Z(\phi)|$ at $\phi=0$ depends on $\theta_0$. Ultimately, under the condition $\omega_0\tau\gg 1$, one needs to compare the function (see Eqs. (\ref{A_ZB_plane})-(\ref{b}) and perform the change of variable $\omega-\omega_0=\Gamma s$)
\begin{equation}
\label{rho}
\rho(\zeta_0,\Gamma,\theta_0)=\frac{\Gamma\tau}{\sqrt{2\pi}}\int_{-\infty}^{\infty} ds\left[\sin^2\left(\frac{\theta_0}{2}\right)e^{\frac{\zeta_0}{1+s^2}}+\cos^2\left(\frac{\theta_0}{2}\right)e^{-\frac{\zeta_0}{1+s^2}}\right]e^{-\frac{\Gamma^2\tau^2s^2}{2}}
\end{equation}
with unity, which corresponds to the no-squeezing case ($\zeta_0=0$). Already an expansion for small values of $\zeta_0$ gives an idea of why larger values of $\theta_0$ imply an enhancement in the spectrum:
\begin{equation}
\rho(\zeta_0,\Gamma,\theta_0)\approx 1-\cos(\theta_0)\sqrt{2\pi}\zeta_0\Gamma\tau.
\end{equation}
However, in our numerical example the amplitude $\zeta_0$ was not much smaller than unity and the expansion for $\zeta_0\ll 1$ does explain the asymmetry in the enhancement for $\theta_0\sim 1$ as compared to the suppression for $\theta_0\ll 1$. We will then assume that $\zeta_0$ is not small, whereas it is realistic to assume that $\Gamma\tau\ll 1$, which is also verified in our numerical example. In this case, one can first rewrite Eq. (\ref{rho}) as
\begin{equation}
\rho(\zeta_0,\Gamma,\theta_0)=1+\frac{\Gamma\tau}{\sqrt{2\pi}}\int_{-\infty}^{\infty}  ds\left[\sin^2\left(\frac{\theta_0}{2}\right)\left(e^{\frac{\zeta_0}{1+s^2}}-1\right)+\cos^2\left(\frac{\theta_0}{2}\right)\left(e^{-\frac{\zeta_0}{1+s^2}}-1\right)\right]e^{-\frac{\Gamma^2\tau^2s^2}{2}}.
\end{equation}
Now, since $\Gamma\tau\ll 1$  and since the functions inside the round brackets vanish for $s\gg \sqrt{\zeta_0}$, by realistically assuming that $\Gamma\tau\ll 1/\sqrt{\zeta_0}$ (which is also fulfilled in our numerical example), the Gaussian function can be approximately set equal to unity. After that, the integral can be evaluated analytically and the result is
\begin{equation}
\begin{split}
&\rho(\zeta_0,\Gamma,\theta_0)\approx 1+\sqrt{\frac{\pi}{2}}\zeta_0\Gamma\tau\\
&\quad\times\left\{\sin^2\left(\frac{\theta_0}{2}\right)e^{\zeta_0/2}\left[\text{I}_0\left(\frac{\zeta_0}{2}\right)-\text{I}_1\left(\frac{\zeta_0}{2}\right)\right]-\cos^2\left(\frac{\theta_0}{2}\right)e^{-\zeta_0/2}\left[\text{I}_0\left(\frac{\zeta_0}{2}\right)+\text{I}_1\left(\frac{\zeta_0}{2}\right)\right]\right\},
\end{split}
\end{equation}
where $\text{I}_{\nu}(\cdot)$ is the modified Bessel function of the first kind of order $\nu$. This expression can be further made more transparent by taking the asymptotic limit of the modified Bessel functions for $\zeta_0\gg 1$ (recall that the strong inequality $\sqrt{\zeta_0}\Gamma\tau\ll 1$ has also to be fulfilled):
\begin{equation}
\rho(\zeta_0,\Gamma,\theta_0)\approx 1+\Gamma\tau\left[\sin^2\left(\frac{\theta_0}{2}\right)\frac{e^{\zeta_0}}{\sqrt{2\zeta_0}}-\cos^2\left(\frac{\theta_0}{2}\right)\sqrt{2\zeta_0}\right].
\end{equation}
This equation clearly provides a qualitative explanation of why one should indeed expect that for $\theta_0$ approaching zero the spectrum will be suppressed but not so strongly as it is enhanced for values of $\theta_0$ increasing towards $\theta_0=\pi$, as it is confirmed numerically. 

\begin{figure}
\begin{center}
\includegraphics[width=0.9\columnwidth]{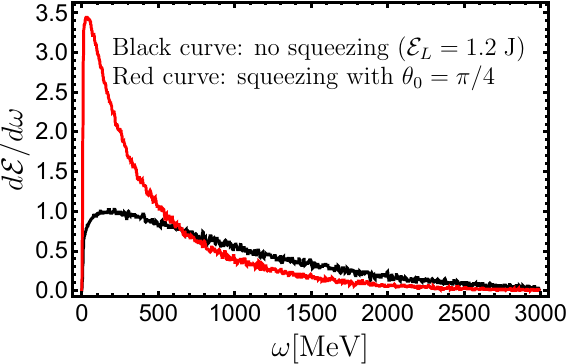}
\end{center}
\caption{Differential energy emitted per unit of photon energy in the presence of a plane wave with and without squeezing with the same pulse energy.}
\label{Spectra_Eq_E}
\end{figure}

An important remark is in order. The process of squeezing involves pumping energy into a crystal where the coherent state to be squeezed passes through. In the process of squeezing energy can be released to or absorbed to the pump. It is not surprising that in the case where we observed enhancement (suppression) of the emission, energy is absorbed from (released to) the pump laser during the squeezing. In fact, in the case of Fig. \ref{Spectra}.a (Fig. \ref{Spectra}.b) the energy of the pulse is $0.3\;\text{J}$ ($1.2\;\text{J}$) as compared to the non-squeezed laser pulse having an energy of about $0.6\;\text{J}$. In the case $\theta_0=\pi/4$ it is useful to compare the spectrum with squeezing with the spectrum in the case of no squeezing but with a laser pulse of energy $1.2\;\text{J}$ (see Fig. \ref{Spectra_Eq_E} and observe that the functions $\xi(\varphi)$ would be identical to those in Fig. \ref{Spectra}.b, with the one in the no-squeezing case only rescaled by the factor $\sqrt{2}$).
The results still show a significant advantage of the squeezing case in the lower part of the spectrum. A single electron emits so many more photons in the squeezing case than in the no-squeezing one (20.8 vs 5.7) that even the average emitted energy is higher in the former case ($1600\;\text{MeV}$ vs $1250\;\text{MeV}$). This shows that a genuine enhancement of radiation is achieved due to squeezing.

\section{Conclusions}
In conclusion, we have studied some effects of squeezing in nonlinear Compton scattering by an intense optical plane-wave field. Since the emitted radiation has frequencies in the gamma domain, we have ignored squeezing effects on the emitted modes. On the contrary, we have assumed that the modes characterizing the background plane wave undergo squeezing although we have realistically ignored the corresponding quantum fluctuations. In fact, within our approach the main effect of squeezing can be described as a frequency modulation of the background plane wave. Our numerical simulations of the collision of an ultrarelativistic electron beam with a frequency-modulated plane wave show that it is possible to enhance or suppress the emission of radiation by an electron as compared with the no-squeezing case. The enhancement is typically more favorably achieved than the suppression even for moderate squeezing parameters, which has also been confirmed analytically. However, the enhancement also requires energy to be pumped into the system during the squeezing process. Nevertheless, a squeezed laser pulse has been shown to induce emission of a larger number of photons and a higher energy than in a non squeezed pulse with the same energy, indicating that squeezing a high-intensity laser field can genuinely enhance the process of radiation.

\begin{acknowledgments}
A.D.P. is partially supported by the U.S. National Science Foundation Mid-scale Research Infrastructure Program under Award No. PHY-2329970. K.Q. is supported by NNSA Grant No. DE-NA0004167 and NSF Grant No. PHY-2308829. 

This material is based upon work supported by the U.S. Department of Energy [National Nuclear Security Administration] University of Rochester ``National Inertial Confinement Fusion Program'' under Award Number DE-NA0004144. 

This report was prepared as an account of work sponsored by an agency of the United States Government. Neither the United States Government nor any agency thereof, nor any of their employees, makes any warranty, express or implied, or assumes any legal liability or responsibility for the accuracy, completeness, or usefulness of any information, apparatus, product, or process disclosed, or represents that its use would not infringe privately owned rights. Reference herein to any specific commercial product, process, or service by trade name, trademark, manufacturer, or otherwise does not necessarily constitute or imply its endorsement, recommendation, or favoring by the United States Government or any agency thereof. The views and opinions of authors expressed herein do not necessarily state or reflect those of the United States Government or any agency thereof.

\end{acknowledgments}


%\bibliography{arXiv,Books,Reviews,Papers_Radiation,Papers_RR,Papers_Various,Homepages,Squeezing}

\end{document}